\documentclass[
aps,%
12pt,%
final,%
notitlepage,%
oneside,%
onecolumn,%
nobibnotes,%
nofootinbib,%
superscriptaddress,%
showpacs,%
centertags]%
{revtex4} 

\def\bb {\begin {eqnarray}}

\def\ee {\end {eqnarray}}

\begin{document}

\title
{In memoriam Yurii Fedorovich Smirnov: \\ 
Some personal reminiscences on a great physicist}

\author{\firstname{Maurice R.}~\surname{Kibler}}


\affiliation{$^a$Universit\'e de Lyon, F--69622, Lyon, France \\ 
             $^b$Universit\'e Claude Bernard Lyon 1, Villeurbanne, France \\
             $^c$CNRS/IN2P3, Institut de Physique Nucl\'eaire de Lyon, France}


\begin{abstract}
Yurii Fedorovich Smirnov (1935-2008) was a famous theoretical physicist. He achieved his career mainly 
at the Institute of Nuclear Physics of Moscow. These notes describe some particular facets of the contributions 
of the late Professor Smirnov in theoretical physics and mathematical physics. They also relate some 
personal reminiscences on Yurii Smirnov in connection with some of his numerous works.
\end{abstract}


\maketitle

\section{Introduction}
Yurii Fedorovich Smirnov passed away in 2008. He was a famous physicist with interests in nuclear physics, 
atomic and molecular physics, condensed matter physics and mathematical physics. More generally, he was at 
the origin of deep achievements around symmetry methods in physics. 

I have not the shoulders to carry the weight of all the fields in which 
Yurii was recognized as a superb researcher. It is enough to say that he 
contributed to many domains of mathematical physics (e.g., finite groups 
embedded in compact or locally compact groups, Lie groups and Lie algebras, 
quantum groups, special functions) and theoretical physics (e.g., nuclear, 
atomic and molecular physics, crystal- and ligand-field theory). Let me 
mention, among other fields, that he obtained alone and with collaborators 
very important results in the theory of clustering in nuclear systems, 
in projection operator techniques, in the theory of 
heavy ion collisions, and in the so-called $J$-matrix formalism for quantum scattering theory. As 
another major contribution, he proposed 
a method, the so-called (e,2e) method, an analog of the (p,2p) method of nuclear 
physics.   

It is not the purpose of these notes to give an extensive list and analyse the numerous papers 
by Yurii. I shall rather focus on some particular facets of his works corresponding to the above 
mentioned fields. I shall also devote a part of these notes to some more personal reminiscences 
related to some of his other fields of interest (crystal-field theory, nonbijective transformations, 
and the use of quantum groups in nuclear physics). 

\section{Yurii Fedorovich Smirnov}

Yurii Fedorovich Smirnov was a Russian physicist. He was born in the city of Il'inskoe (Yaroslavl' 
region, Russia) in 1935. He graduated from Moscow State University. Then, he completed his 
Doctorate thesis at the same university under the guidance of Yurii M. Shirokov and 
benefited from fruitful contacts with other distinguished physicists as for instance Yakov A. Smorodinsky. He 
achieved his career in the (Skobeltsyn) Institute of Nuclear Physics and in the Physics Department of (Lomonosov) 
Moscow State University with many stays abroad. The 
last fifteen years of his life were shared between Moscow and Mexico city where he was a visiting Professor and got 
a Professor position successively at the {\it Instituto de F\'\i sica} and, then, at the {\it Instituto de 
Ciencias Nucleares} of the U.N.A.M. (he spent almost 11 years in Mexico). He received prestigious awards: 
the K.D. Sinel'nikov Prize of the Ukrainian Academy of Sciences in 1982 and the M.V. Lomonosov Prize in 2002. He was a 
member of the Academy of Sciences of Mexico. 

Yurii Smirnov signed and/or co-signed eleven books and more than 250 scientific articles (only a few part of 
them shall be mentionned here). He translated into Russian several scientific books. As an example, he translated 
a book on the harmonic oscillator written by Marcos Moshinsky in 1969, precisely the book which was a starting 
point for their common book, on the same subject, published in 1996 \cite{MosSmi}. He was a member of the editorial 
board of several journals and a councillor of the scientific councils of the Skobeltsyn Institute of Nuclear 
Physics and of the Chemistry Department of Moscow State University as well as of the Institute for 
Theoretical and Experimental Physics (ITEP) in Moscow. 

\section{A first seminal work}

Following the pioneer work of Talmi \cite{Talmi1952}, Marcos Moshinsky and Yurii F. Smirnov were independently 
interested in the description of pairs of nucleons in a harmonic-oscillator potential. In 1959, Moshinsky developed a 
formalism to connect the wave functions in two different coordinate systems for two particles (with identical masses) 
in a harmonic-oscillator potential \cite{Mos1959}. In this formalism, any two-particle wave function 
$|n_1 \ell_1, n_2 \ell_2, \lambda \mu \rangle$, expressed in coordinates with respect to the 
origin of the harmonic-oscillator potential, is a linear combination of wave functions $|n \ell, N L, \lambda \mu \rangle$, 
expressed in relative and centre-of-mass coordinates of the two particles. The so-called 
transformation brackets $\langle n \ell, N L, \lambda | n_1 \ell_1, n_2 \ell_2, \lambda \rangle$ make it possible to pass 
from one coordinate system to the other. Moshinsky gave an explicit expression of these coefficients in the case 
$n_1 = n_2 = 0$ and derived recurrence relations that can be used to obtain the coefficients for 
$n_1 \not= 0$ and $n_2 \not= 0$ from those for $n_1 = n_2 = 0$ \cite{Mos1959}. Along this vein, Brody and Moshinsky 
published extensive tables of transformation brackets \cite{BroMos}. At the end of the fiftees, Smirnov worked out a 
parent problem, viz. the calculation of the Talmi coefficients for unequal mass nucleons, and he gave a solution for 
the case $n_1 \not= 0$ and $n_2 \not= 0$ \cite{Smi19611962}. (Indeed, the transformation brackets and the Talmi 
coefficients are connected via a double Clebsch-Gordan transformation.) The coefficients 
$\langle n \ell, N L, \lambda | n_1 \ell_1, n_2 \ell_2, \lambda \rangle$, called {\it transformation brackets} by Moshinsky 
and {\it total Talmi coefficients} by Smirnov, are now referred to as Moshinsky-Smirnov coefficients. In this respect, both the 
Moshinsky-Smirnov coefficients and the Talmi coefficients were revisited at the end of the seventies in terms of 
generating functions in the framework of the approaches by J.S. Schwinger and by V. Bargmann of the 
harmonic-oscillator bases \cite{Mehdi1980}. (The work by M. Hage Hassan \cite{Mehdi1980}, who prepared his 
State Doctorate thesis at the {\it Institut de Physique Nucl\'eaire de Lyon}, constitutes a very deep and original 
approach to the Talmi coefficients and Moshinsky-Smirnov coefficients.) It should be noted that the transformation brackets 
or Moshinsky-Smirnov coefficients are also of importance for atoms and molecules as shown by Marcos and Yurii in their book \cite{MosSmi}
written during the time Yurii was an invited professor at the {\it Instituto de F\'\i sica} of the 
{\it Universidad Nacional Aut\'onoma de M\'exico}. 

The harmonic oscillator is a central ingredient in numerous works by Smirnov and Moshinsky. Many applications 
of the nonrelativistic and relativistic harmonic oscillators to modern physics (from molecules, atoms, and nuclei 
to quarks) were pedagogically exposed in the book by Marcos and Yurii \cite{MosSmi} with a special attention paid 
to the many-body problem (in the Hartree-Fock approximation and from the point of view of unitary symmetry), the 
nuclear collective motion, and the interacting boson model. Their common interests were not limited to applications  
based on the harmonic oscillator. Let us simply mention that both of them were interested in group theoretical 
methods and symmetry methods in physics and also contributed to several fields of mathematical physics as, for 
instance, state labelling problem, special functions, generating functions, and nonbijective canonical 
transformations. We shall continue in Section IV with some contributions of Yurii mainly in nuclear physics 
with some excursions in some other domains.  

\section{Contributions in nuclear physics and related areas}

\subsection{Nuclear structure models}

Yu.F. Smirnov and some of his collaborators contributed significantly to the theory of 
clustering in nuclei (see the monographs \cite{N, NeNRT}). His researches 
in this domain started, as seen above, with the derivation of 
transformation brackets \cite{Smi19611962}. Then, it was necessary to 
elaborate a formalism for deriving many-nucleon fractional parentage coefficients (FPCs) including one, 
two, and sometimes three shells. This problem was solved by Yu.F. Smirnov and co-workers via a 
factorization of each FPC into a space part and a spin-isospin one \cite{KuM}. The coincidence of the 
spin-isospin part of the FPCs with the isoscalar factor part of the Clebsh-Gordan coefficients of the group 
$SU(4)$ was demonstrated.

Calculation technique of cluster characteristics of oscillator shell-model systems
(spectroscopic amplitudes, factors, etc.) as a whole was first developed in 
 \cite{BNY, Chl}. For 1$p$-shell nuclei, a complete determination of these 
characteristics was presented in \cite{T1}.

For lightest nuclei Yu.F. Smirnov with co-workers built 
the translationally-invariant shell model which is free 
of nonphysical oscillations of the center of mass of a 
nucleus \cite{KShE}. 

Cluster model was proposed by Yu.F. Smirnov with co-authors for the description of
electromagnetic form-factors of lightest nuclei \cite{KudKN}.

The formalism of FPCs for many-boson systems, as nuclei
described by the interacting boson model in the $SU(3)$ limit, was built by Yu.F. Smirnov and
co-authors in \cite{GKh}.

In connection with shell model calculations, Yu.F. Smirnov and K.V. Shitikova \cite{Sh} gave 
a significant contribution to the hyperspherical function method. In the same direction,  he 
also developed some group theoretical aspects of the generalized hyperspherical function method 
\cite{FO}. As a further important contribution in the field of mathematical physics, Smirnov  
developed, in collaboration with R.M. Asherova and V.N. Tolstoy, the method of extremal projection operators
for simple Lie groups \cite{Tolstoy1} and for semisimple complex Lie algebras \cite{Tolstoy2} (this 
powerful and universal method was generalized by V.N. Tolstoy to all finite-dimensional contragredient
superalgebras, to affine Kac-Moody algebras and superalgebras, and finally to the $q$-deformations of these 
algebras and superalgebras). Yu.F. Smirnov and collaborators applied the extremal projection operator method
to several groups \cite{Tolstoy3}-\cite{Tolstoy10}; in particular, he elaborated a formalism for the Clebsch-Gordan 
coefficients and the 3$nj$--symbols of $SU(3)$ and of the $q$-deformations of $SU(2)$ and $SU(3)$.

\subsection{Theory of nuclear reactions}

The theory of cluster quasi-elastic knock-out nuclear reactions was developed by
Yu.F. Smirnov and co-authors in \cite{BZN}. The role of cluster de-excitation in this 
process was demonstrated in \cite{GINT}. 

Yu.F. Smirnov and Yu.M. Tchuvil'sky pointed out by making use of group-theory methods a
general property of heavy projectile -- heavy target nucleus channel, namely, systems 
exhibiting great
distinctions between the wavefunction of the channel and the wavefunction of the
respective compound. As a result, the concept of the co-called structural forbidness of
heavy fragmentation (and fusion) was introduced into the theory of heavy ion collisions
 \cite{T2}.

Calculation methods concerning the theory of alpha decay of giant isoscalar quadrupole
resonances of nuclei were presented in the paper \cite{T3}.

\subsection{Quark degrees of freedom of nuclear systems}

Independently of V.A. Matveev and P. Sorba \cite{Sorba} -- the authors of the
concept of hidden color of the six-quark system -- Yu.F. Smirnov and Yu.M. Tchuvil'sky
demonstrated that if the deuteron would consist of two three-quark nucleons, 
then its wavefunction should contain components with hidden nucleon isobars ($\Delta$, etc.) 
\cite{T4}. The method of calculation of the statistical weights of these components in
the deuteron wavefunction was worked out in \cite{T4}.

Yu.F. Smirnov was an active participant of the team who developed methods of calculation
relative to the dynamics of the six-quark system \cite{OT}. (The paper \cite{OT} contains 
a proof of the connection between the isoscalar factors of the Clebsch-Gordan coefficients 
of the unitary group $SU(n)$ and the Clebsch-Gordan coefficients of the companion permutation 
group $S_n$.) This team also emphasized the role of color-magnetic attraction and as a 
consequence of quark configuration mixing in this system \cite{ONT1}. 

\subsection{Composite particle interaction}

Yu.F. Smirnov and co-authors proposed an approach for treating the interaction between
composite particles through the use of deep attractive potentials with redundant states 
\cite{KuN}. He was the principal researcher of the group which made the first calculations of the
ground state of the three-body (3$\alpha$) system with an interaction of such a type \cite{ONT2}.

\subsection{The $J$-matrix formalism}

The $J$-matrix formalism, a formalism utilizing $L^2$ bases in quantum scattering theory, 
was suggested in the mid-seventies by E.J. Heller, H.A. Yamani, W.P. Reinhardt, and 
L. Fishman, people coming from the atomic physics community. The same formalism utilizing the 
oscillator basis was independently rediscovered in 1980 by G.F. Filippov, a good friend 
of Yurii. Then, Yu.F. Smirnov become very interested in this new quantum scattering formalism 
and performed a lot of works by developing and using it in various applications. In particular, 
(with Yu.I. Nechaev) he developed exact mathematical grounds of the $J$-matrix
approach with the help of an oscillator basis \cite{NeSm}, 
studied (together with J.M. Bang, A.I. Mazur, A.M. Shirokov, and S.A. Zaytsev) 
its relations with the well-known $R$-matrix and $P$-matrix formalisms in scattering 
theory \cite{118}, extended 
(together with A.M. Shirokov and S.A. Zaitsev) the $J$-matrix formalism to the case of the so-called 
true few-body scattering \cite{TMF} and to the case of relativistic systems described by the 
Dirac equation. The first applications of the
$J$-matrix formalism (in the oscillator basis) to the nuclear shell model
was presented in a series of papers, published by Yu.F. Smirnov 
with A.I. Mazur and V.A. Knyr, devoted to the hypernucleus production
reactions. Another application of this formalism, an application to systems decaying via three-body channels,
was published by Yu.F. Smirnov with T.Ya. Mikhelashvili and A.M. Shirokov \cite{23}. Later Yu.F. Smirnov 
with Yu.A. Lurie and A.M. Shirokov studied in detail, by means of this
approach, three-body decays of loosely-bound $^{11}$Li and $^6$He
nuclei within cluster models. As further interesting extensions of this
approach, the three-body $J$-matrix
formalism was used to calculate the three-body $S$-matrix and a direct
numerical calculation of the $S$-matrix poles made it possible to
improve essentially variational calculation results for the binding
energies, ground state rms radii, electromagnetic transitions, etc. It
is worth noting that the first $J$-matrix calculation of three-body
$S$-matrix poles was performed by Yurii Smirnov together with
A.M. Shirokov and L.Ya. Stotland in the studies of atomic He and H$^-$ ion.
A review of some of the results obtained using the $J$-matrix approach
by Yurii Smirnov and collaborators can be found in \cite{Cocoyoc}.

The $J$-matrix formalism is based on the solutions of three-term
recurrence relations. This brought naturally Yurii Smirnov to study 
general properties of eigenenergies and eigenvectors for three-term
recurrence relations (equivalent to second-order finite difference
equations). Together with P.A. Braun, A.M. Shirokov, and S.K. Suslov he
developed an approximate method for analyzing the spectrum structure
and eigenvector properties of three-term recurrence relations by
replacing them by second-order ordinary differential equations of the 
Schr\"odinger type, a replacement which makes it possible to develop quantum
intuition. This approach was used by Yu.F. Smirnov and collaborators in
various problems, as for instance in the analysis of general properties of Clebsch--Gordan and Racah coefficients 
for $SU(2)$ and $SU(1,1)$ groups \cite{Suslov} as well as in the study of level clustering in the high-$J$
spectra of non-rigid spherical top molecules \cite{Braun} 
and of partly-filled shell ions in crystalline fields \cite{Svir}. Interesting results were also 
obtained by Yurii Smirnov and collaborators in the study of exact
solutions of three-term recurrence relations. In particular, together
with A.M. Shirokov and N.A. Smirnova, he discovered a parameter symmetry
of the interaction boson model \cite{PS} 
which was studied by him later (with O. Costa\~nos, A. Frank, A.M. Shirokov, and N.A. Smirnova) in more
complicated versions of the interacting boson model and in some other
algebraic models.

\subsection{The (e,2e) method}

In collaboration with V.G. Neudatchin, in 1967--1969, Yu.F. Smirnov proposed a 
new method for the experimental investigation of the electronic structure of atoms, 
molecules, and solids. This method, based on the measurement of the process of 
quasi-elastic knock-out of electron by high-energy (in atomic scale) electron 
with coincidences registration, is called the (e,2e) method. It is the natural 
analog of the (p,2p) method popular in nuclear physics. The proposed method was 
successfully applied in many laboratories around the world and provided data on 
the structure of many-electron systems (see for example \cite{NPS} for a review 
about this major contribution).

\section{Some personal reminiscences}

My first contact with the work of Yurii Smirnov goes back to 1978 when my colleague 
J. Patera showed me, on the occasion of a NATO Advanced Study Institute organised in Canada by J.C. Donini, 
a beautiful book written by D.T. Sviridov and Yu.F. Smirnov \cite{Smirnov1}. This book 
dealt with the spectroscopy of $d^N$ ions in inhomogeneous electric fields (a part of a 
disciplinary domain known as crystal- and ligand-field theory in condensed matter physics and 
explored via the theory of level splitting from a theoretical point of view). In 1979, B.I. Zhilinski\u\i , 
while visiting Dijon and Lyon in France in the framework of an exchange programme between 
USSR and France, provided me with another interesting book, dealing with $f^N$ ions in 
crystalline fields, written by D.T. Sviridov, Yu.F. Smirnov and V.N. Tolstoy \cite{Smirnov2}. At 
that time, the references for mathematical aspects of crystal- and ligand-field theory were based 
on works by Y. Tanabe, S. Sugano, and H. Kamimura from Japan \cite{Tanabe}, J.S. Griffith from 
England \cite{Griffith}, and Tang Au-chin and his collaborators from China \cite{Tang} (see also 
some contributions by the present author \cite{KibJMSIJQC}). The two above-mentioned books 
by Smirnov and his colleagues shed some new light on the mathematical analysis of spectroscopic 
and magnetic properties of partly filled shell ions in molecular and crystal surroundings. In 
particular, special emphasis was put on the derivation of the Wigner-Racah algebra of a finite 
group of molecular and crystallographic interest from that of the group $SO(3) \sim SU(2)/Z_2$. 

My second (indirect) contact with Yurii is related to an invitation to participate in the fifth workshop on 
{\it Symmetry Methods in Physics} organized by Yurii F. Smirnov and Raya M. Asherova in Obninsk in July 1991. 
Unfortunately, I did not get my visa on time so that my participation was reduced to a paper in the proceedings 
of the workshop edited by Smirnov and Asherova \cite{KiblerObninsk}.  

In the beginning of the 1990's, I had a chance to get in touch with another facet of Yurii's work. In 1989, 
a Russian speaking student from Switzerland, C. Campigotto, spent one year in the group of Prof.~Smirnov. He 
started working on the so-called Kustaanheimo-Stiefel transformation, an ${\bf R}^4 \to {\bf R}^3$ 
transformation associated with the Hopf fibration ${S}^3 \to {S}^2$ with compact fiber $S^1$. Such a nonbijective transformation 
makes it possible to connect the Kepler-Coulomb system in ${\bf R}^3$ to the isotropic harmonic oscillator 
in ${\bf R}^4$. (More precisely, the Kustaanheimo-Stiefel transformation allows to pass from a 
four-dimensional harmonic oscillator subjected to a constrain to the three-dimensional hydrogen atom, 
see for instance \cite{KibNeg}). Then, Campigotto (well-prepared by Smirnov and his team, especially A.M. Shirokov and V.N. Tolstoy) 
came to Lyon to prepare a French doctorate thesis \cite{Campigotto}. He defended his thesis in 1993 with G.S. Pogosyan 
(representing Yu.F. Smirnov) as a member of the jury. 

A fourth opportunity to get involved with Yurii came from our mutual interest in quantum groups 
and in nuclear and atomic spectroscopy. I meet him for the first time in Dubna in 1992. We then 
started a collaboration (partly with R.M. Asherova) on $q$- and $qp$-boson calculus in the 
framework of Hopf algebras associated with the Lie algebras $su(2)$ and $su(1,1)$ \cite{KibSmiCamAsh}. In addition, 
we pursued a group theoretical study of the Coulomb energy averaged over the $n \ell^N$--atomic 
states with a definite spin \cite{IJQC}. We also had fruitful exchanges in nuclear physics. Indeed, Prof.~Smirnov 
and his colleagues D. Bonatsos (from Greece), S.B. Drenska, P.P. Raychev and R.P. Roussev (all from Bulgaria) 
developed a model based on a one-parameter deformation of $SU(2)$ for dealing with rotational 
bands of deformed nuclei and rotational spectra of molecules \cite{Bonatsos} (see also \cite{Georgieva}). Along 
the same line, a student of mine, R. Barbier, developed in his thesis a 
two-parameter deformation of $SU(2)$ with application to superdeformed nuclei in mass region $A \sim 130-150$  
and $A \sim 190$ \cite{Barbier}. It was a real pleasure to receive Yurii in Lyon on the 
occasion of the defence of the Barbier thesis in 1995. Indeed, from 1992 to 1995, Yurii made four stays in 
Lyon (one with his wife Rita and one with his daughter Tatyana) and we jointly participated in several meetings, 
one in Clausthal in Germany (organised by H.-D. Doebner, V.K. Dobrev, and A.G. Ushveridze) and two in Bregenz in 
Austria (organised by B. Gruber and M. Ramek).

\section{Closing} 
Yurii had many students (some of them are now famous physicists), many collaborators in 
his country and abroad, and had an influence on many scientists. He was 
also an exceptional teacher. It was very pleasant, profitable and inspiring 
to be taught by Prof.~Smirnov and/or to discuss with him. I personally greatly 
benefited from discussions with Yurii Smirnov.

Yurii Fedorovich Smirnov will remain an example for many of us. We will 
remember the exceptional qualities of the man as a scientist, as a teacher 
and as a generous person. 

Yurii, we shall not forget you.

\section*{Acknowledgments}
Part of these notes were presented at the 13th International Conference on Symmetry Methods in Physics 
(SYMPHYS-XIII) organized in memory of Prof.~Yurii Fedorovich Smirnov by the Bogoliubov Laboratory of 
Theoretical Physics of the Joint Institute for Nuclear Research and the International Center for 
Advanced Studies at Yerevan State University (the conference was held in Dubna, Russia, 6-9 July 2009). I
acknowledge the Organizing Committee of SYMPHYS-XIII, especially George S. Pogosyan, for their kind invitation to participate to 
this interesting conference. The part not presented at SYMPHYS-XIII were written on the basis of notes 
communicated by Vladimir G. Neudatchin, Andrey M. Shirokov, Yurii M. Tchuvil'sky, and Valeriy N. Tolstoy. I am 
very much indebted to them for their invaluable help in the preparation of this paper. Thanks are also due to 
Corrado Campigotto, Mehdi Hage Hassan, and Rutwig Campoamor-Stursberg for useful correspondence.


\begin{thebibliography}{99}

\bibitem{MosSmi}
M. Moshinsky and Yu.F. Smirnov, 
{\it The Harmonic Oscillator in Modern Physics} 
(Harwood Academic Publishers, Amsterdam, 1996).

\bibitem{Talmi1952} 
I. Talmi, 
Helv. Phys. Acta {\bf 25}, 185 (1952).  

\bibitem{Mos1959} 
M. Moshinsky, 
Nucl. Phys. {\bf 13}, 104 (1959). 

\bibitem{BroMos} 
T.A. Brody and M. Moshinsky, 
{\it Tables of Transformation Brackets for Nuclear Shell-Model Calculations}
(Universidad Nacional Aut\'onoma de M\'exico, Mexico city, 1960 and 
Gordon and Breach Science Publishers, New York, 1967).

\bibitem{Smi19611962} 
Yu.F. Smirnov, 
Nucl. Phys. {\bf 27}, 177 (1961); {\bf 39}, 346 (1962). 

\bibitem{Mehdi1980} 
M. Hage Hassan,
J. Phys. A: Math. Gen. {\bf 13}, 1903 (1980). 

\bibitem{N} 
V.G. Neudatchin and Yu.F. Smirnov, 
{\it Nucleon Clusters in Light Nuclei} (Nauka, Moscow, 1969) (in Russian).

\bibitem{NeNRT} 
O.F. Nemets, V.G. Neudatchin, A.T. Rudchik, Yu.F. Smirnov, and Yu.M. Tchuvil'sky, 
{\it Nucleon Clusters in Atomic Nuclei and Many-Nucleon Transfer Reactions} (Naukova Dumka, Kiev, 1988) (in Russian).

\bibitem{KuM} 
V.I. Kukulin, L. Majling, and Yu.F. Smirnov, 
Nucl. Phys. A {\bf 103}, 681 (1968).

\bibitem{BNY} 
V.V. Balashov, V.G. Neudatchin, Yu.F. Smirnov, and N.P. Yudin, 
ZhETP. {\bf 37}, 1385 (1959) (in Russian).

\bibitem{Chl} 
Yu.F. Smirnov and D. Chlebovska, 
Nucl. Phys. {\bf 26}, 306 (1961).

\bibitem{T1} 
Yu.F. Smirnov and Yu.M. Tchuvil'sky, 
Phys. Rev. C {\bf 15}, 84 (1977).

\bibitem{KShE} 
I.V. Kurdyumov, Yu.F. Smirnov, K.V. Shitikova, and S.H. El-Samarae, 
Nucl. Phys. A {\bf 145}, 593 (1970).

\bibitem{KudKN} 
Yu.A. Kudeyarov, I.V. Kurdyumov, V.G. Neudatchin, and Yu.F. Smirnov, 
Nucl. Phys. A {\bf 163}, 316 (1971).

\bibitem{GKh} 
I.S. Guseva, Yu.F. Smirnov, and Yu.I. Kharitonov, 
Izv. AN SSSR, Ser. fiz. {\bf 49}, 37 (1985) (in Russian).

\bibitem{Sh} 
Yu.F. Smirnov and K.V. Shitikova, 
Physics of Elementary Particles and Atomic Nuclei {\bf 8}, 847 (1977) (in Russian).

\bibitem{FO} 
G.F. Filippov, V.I. Ovcharenko, and Yu.F. Smirnov, 
{\it Microskopic Theory of Collective Excitations of Atomic Nuclei} (Naukova Dumka, Kiev, 1981) (in Russian).

\bibitem{Tolstoy1} 
R.M. Asherova, Yu.F. Smirnov, and V.N. Tolsto\u\i, 
Teoret. Mat. Fiz. {\bf 8}, 255 (1971). 

\bibitem{Tolstoy2}
R.M. Asherova, Yu.F. Smirnov, and V.N. Tolsto\u\i, 
Mat. Zametki {\bf 26}, 15 (1979).


\bibitem{Tolstoy3}
R.M. Asherova, Yu.F. Smirnov, and V.N. Tolsto\u\i, 
Teoret. Mat. Fiz. {\bf 15}, 107 (1973).

\bibitem{Tolstoy4}
Z. Pluha\v r, Yu.F. Smirnov, and V.N. Tolstoy, 
J. Phys. A: Math. Gen. {\bf 19}, 21 (1986).

\bibitem{Tolstoy5}
Yu.F. Smirnov, V.N. Tolsto\u\i, and Yu.I. Kharitonov, 
Yad. Fiz. {\bf 53}, 959 (1991) [Soviet J. Nucl. Phys. {\bf 53}, 593 (1991)].

\bibitem{Tolstoy6}
Yu.F. Smirnov, V.N. Tolsto\u\i, and Yu.I. Kharitonov, 
Yad. Fiz. {\bf 53}, 1746 (1991) [Soviet J. Nucl. Phys. {\bf 53}, 1068 (1991)].

\bibitem{Tolstoy7}
Yu.F. Smirnov, V.N. Tolsto\u\i, and Yu.I. Kharitonov, 
Yad. Fiz. {\bf 55}, 2863 (1992) [Soviet J. Nucl. Phys. {\bf 55}, 1599 (1992)]. 

\bibitem{Tolstoy8}
Yu.F. Smirnov, V.N. Tolsto\u\i, and Yu.I. Kharitonov, 
Yad. Fiz. {\bf 56}, 223 (1993) [Soviet J. Nucl. Phys. {\bf 56}, 690 (1993)]. 

\bibitem{Tolstoy9}
R.M. Asherova, Yu.F. Smirnov, and V.N. Tolstoy, 
Yad. Fiz. {\bf 59}, 1859 (1996) [Phys. Atomic Nuclei {\bf 59}, 1795 (1996)]. 

\bibitem{Tolstoy10}
R.M. Asherova, Yu.F. Smirnov, and V.N. Tolstoy,  
Yad. Fiz. {\bf 64}, 2170 (2001) 
[Phys. Atomic Nuclei {\bf 64}, 2080 (2001)].

\bibitem{BZN} 
P. Beregi, N.S. Zelenskaya, V.G. Neudatchin, and Yu.F. Smirnov, 
Nucl. Phys. A {\bf 66}, 513 (1965).

\bibitem{GINT} 
N.F. Golovanova, I.M. Il'in, V.G. Neudatchin, Yu.F. Smirnov, and Yu.M. Tchuvil'sky, 
Nucl. Phys. A {\bf 262}, 444 (1976).

\bibitem{T2} 
Yu.F. Smirnov and Yu.M. Tchuvil'sky, 
Phys. Lett. B {\bf 134}, 25 (1984).

\bibitem{T3} 
Yu.F. Smirnov and Yu.M. Tchuvil'sky, 
Izv. AN SSSR, Ser. fiz. {\bf 47}, 151 (1983) (in Russian).

\bibitem{Sorba} 
V.A. Matveev and P. Sorba, 
Lett. Nuovo Cimento {\bf 20}, 435 (1977).

\bibitem{T4}
Yu.F. Smirnov and Yu.M. Tchuvil'sky, 
J. Phys. G: Nucl. Part. Phys. {\bf 4}, L1 (1978).

\bibitem{OT} 
I.T. Obukhovsky, Yu.F. Smirnov, and Yu.M. Tchuvil'sky, 
J. Phys. A: Math. Gen. {\bf 15}, 7 (1982).

\bibitem{ONT1}    
I.T. Obukhovsky, V.G. Neudatchin, Yu.F. Smirnov, and Yu.M. Tchuvil'sky, 
Phys. Lett. B. {\bf 88}, 231 (1979).

\bibitem{KuN}
V.I. Kukulin, V.G. Neudatchin, and Yu.F. Smirnov, 
Nucl. Phys. A {\bf 245}, 429 (1975).

\bibitem{ONT2}  
Yu.F. Smirnov, I.T. Obukhovsky,  Yu.M. Tchuvil'sky, and V.G. Neudatchin, 
Nucl. Phys. A {\bf 235}, 289 (1974).

\bibitem{NeSm} 
Yu.F. Smirnov and Yu.I. Nechaev, 
Kinam {\bf 4}, 445 (1982);
Yu.I. Nechaev and Yu.F. Smirnov, 
Yad. Fiz. {\bf 35}, 1385 (1982) [Soviet J. Nucl. Phys. {\bf 35}, 808 (1982)].

\bibitem{118} 
J.M. Bang, A.I. Mazur, A.M. Shirokov, Yu.F. Smirnov, and S.A. Zaytsev, 
Ann. Phys. (N.Y.) {\bf 280}, 299 (2000).  

\bibitem{TMF} 
S.A. Zaitsev, Y.F. Smirnov, and A.M. Shirokov, 
Teoret. Mat. Fiz. {\bf 117}, 1291 (1988) [Theor. Math. Phys. {\bf 117}, 1291 (1988)].  

\bibitem{23} 
T.Ya. Mikhelashvili, A.M. Shirokov, and Yu.F. Smirnov,
J. Phys. G: Nucl. Part. Phys. {\bf 16}, 1241 (1990). 

\bibitem{Cocoyoc} 
Yu.F. Smirnov, A.M. Shirokov, Yu.A. Lurie, and S.A. Zaytsev, 
Harmonic oscillator representation in the theory of scattering and nuclear reactions, 
in: {\it Second Int. Workshop on Harmonic Oscillators. Cocoyoc, Morelos, Mexico,   
March 23--25, 1994}, eds. D. Han and K.B. Wolf (NASA, Washington, 1995).

\bibitem{Suslov} 
Yu.F. Smirnov, S.K. Suslov, and A.M. Shirokov, 
J. Phys. A: Math. Gen. {\bf 17}, 2157 (1984).

\bibitem{Braun} 
P.A. Braun, A.M. Shirokov, and Yu.F. Smirnov, 
Molec. Phys. {\bf 56}, 573 (1985).

\bibitem{Svir} 
D.T. Sviridov, Yu.F. Smirnov, and A.M. Shirokov,  
Dokl. Akad. Nauk SSSR {\bf 317}, 98 (1991) [Soviet Phys. Dokl. {\bf  36}, 229 (1991)]. See also: 
A.M. Shirokov and Yu.F. Smirnov, 
J. Phys. A: Math. Gen. {\bf 24}, 2961 (1991).

\bibitem{PS} 
A.M. Shirokov, N.A. Smirnova, and Yu.F. Smirnov,
Phys. Lett. B {\bf 434}, 237 (1998).

\bibitem{NPS} 
V.G. Neudatchin, Yu.V. Popov, and Yu.F. Smirnov, 
Uspekhi Fiz. Nauk {\bf 169}, 1111 (1999).

\bibitem{Smirnov1}
D.T. Sviridov and Yu.F. Smirnov, 
{\it Teoriya Opticheskikh Spektrov Perekhodnykh Metallov} (Izd. Nauka, Moscow, 1977). See also: 
D.T. Sviridov and Yu.F. Smirnov, Soviet. Phys. Doklady {\bf 13}, 565 (1968); 
C.V. Vonsovski, C.V. Grimailov, V.I. Tcherepanov, A.N. Meng, D.T. Sviridov, Yu.F. Smirnov, and A.E. Nikiforov, 
{\it Crystal-Field Theory and Optical Spectra of Partly Filled $d$ Shell Transition Ions} (Nauka, Moscow, 1969), in Russian; 
D.T. Sviridov, R.K. Sviridova, and Yu.F. Smirnov, {\it Optical Spectra of Transition-Metal Ions in Crystal} (Nauka, Moscow, 1976), in Russian.

\bibitem{Smirnov2}
D.T. Sviridov, Yu.F. Smirnov, and V.N. Tolstoy, 
{\it Spektroskopiya Kristallov} (Akad. Nauk SSSR, Moscow, 1975). 

\bibitem{Tanabe}
Y. Tanabe and S. Sugano, J. Phys. Soc. Japan {\bf 9}, 753 (1954); {\bf 9}, 766 (1954); {\bf 11}, 864 (1956).
Y. Tanabe and H. Kamimura, J. Phys. Soc. Japan {\bf 13}, 394 (1958). 
S. Sugano and Y. Tanabe, J. Phys. Soc. Japan {\bf 13}, 880 (1958). See also: 
S. Sugano, Y. Tanabe, and H. Kamimura, 
{\it Multiplets of Transition-Metal Ions in Crystals} (Academic Press, New York, 1970).

\bibitem{Griffith}	
J.S. Griffith, Molec. Phys. {\bf 3}, 79 (1960); {\bf 3}, 285 (1960); {\bf 3}, 457 (1960); {\bf 3}, 477 (1960).  
See also:
J.S. Griffith, 
{\it The Theory of Transition-Metal Ions} (Cambridge Univ. Press, Cambridge, 1961);	
{\it The Irreducible Tensor Method for Molecular Symmetry Groups} (Prentice-Hall, Englewood Cliffs, 1962). 

\bibitem{Tang}
Tang Au-chin, Sun Chia-chung, Kiang Yuan-sun, Deng Zung-hau, Liu Jo-chuang, Chang Chian-er, Yan Go-sen, Goo Zien, and Tai Shu-shan, Sci. Sinica (Peking) {\bf 15}, 610 (1966). See also: 
Tang Au-chin, Sun Chia-chung, Kiang Yuan-sun, Deng Zung-hau, Liu Jo-chuang, Chang Chain-er, Yan Guo-sen, Goo Zien, and Tai Shu-shan, 
{\it Theoretical Method of the Ligand Field Theory} (Science Press, Peking, 1979). 

\bibitem{KibJMSIJQC}
M. Kibler, J. Molec. Spectrosc. {\bf 26}, 111 (1968); 
           Int. J. Quantum Chem. {\bf 3}, 795 (1969); 
           C.R. Acad. Sci. (Paris), Ser. B {\bf 268}, 1221 (1969). See also: 
           M.R. Kibler, Group theory around ligand field theory, in: {\it Group Theoretical Methods in Physics}, 
           eds. R.T. Sharp and B. Kolman (Academic Press, New York, 1977); Finite symmetry adaptation in spectroscopy, in: 
           {\it Recent Advances in Group Theory and Their Application to Spectroscopy}, ed. J.C. Donini (Plenum Press, New York, 1979). 

\bibitem{KiblerObninsk}
M. Kibler and M. Daoud, Symmetry adaptation and two-photon spectroscopy of ions in molecular or 
solid-state finite symmetry, in: {\it Symmetry Methods in Physics}, eds. 
Yu.F. Smirnov and R.M. Asherova (Russian Federation Ministry of Atomic Energy -- Institute of Physics and Power Engineering, Obninsk, 1992).

\bibitem{KibNeg}
M. Kibler and T. N\'egadi,
Lett. Nuovo Cimento {\bf 37}, 225 (1983);  
J. Phys. A: Math. Gen. {\bf 16}, 4265 (1983);  
Phys. Rev. A {\bf 29}, 2891 (1984).

\bibitem{Campigotto}
C. Campigotto, Doctorate thesis, Universit\'e Lyon 1, France, 1993. The thesis was partly published in:  
C. Campigotto and Yu.F. Smirnov, Helv. Phys. Acta {\bf 64}, 48 (1991);   
Gh.E. Dr\u{a}g\u{a}nescu, C. Campigotto, and M. Kibler, Phys. Lett. A {\bf 170}, 339 (1992);
M. Kibler and C. Campigotto, Int. J. Quantum Chem. {\bf 45}, 209 (1993); 
Phys. Lett. A {\bf 181}, 1 (1993); 
C. Campigotto, Yu.F. Smirnov, and S.G. Enikeev, 
J. Comp. Appl. Math. {\bf 57}, 87 (1995); 
Yu.F. Smirnov and C. Campigotto, 
J. Comp. Appl. Math. {\bf 164--165}, 643 (2004). 

\bibitem{KibSmiCamAsh}
Yu.F. Smirnov and M.R. Kibler, 
Some aspects of $q$-boson calculus, in: 
{\it Symmetries in Science VI: From the Rotation Group to Quantum Algebras},  
ed. B. Gruber (Plenum Press, New York, 1993).
M. Kibler, C. Campigotto, and Yu.F. Smirnov, 
Recursion relations for Clebsch-Gordan coefficients of $U_q(su_2)$ and $U_q(su_{1,1})$, in: 
{\it Symmetry Methods in Physics}, eds.
A.N. Sissakian, G.S. Pogosyan, and S.I. Vinitsky (Joint Institute for Nuclear Research, Dubna, 1994).
M.R. Kibler, R.M. Asherova, and Yu.F. Smirnov, 
Some aspects of $q$- and $qp$-boson calculus, in:
{\it Symmetries in Science VIII}, ed. B. Gruber (Plenum Press, New York, 1995).

\bibitem{IJQC}
M. Kibler and Yu.F. Smirnov, 
Int. J. Quantum Chem. {\bf 53}, 495 (1995). 

\bibitem{Bonatsos}
D. Bonatsos, S.B. Drenska, P.P. Raychev, R.P. Roussev, and Yu.F. Smirnov, 
J. Phys. G: Nucl. Part. Phys. {\bf 17}, L67 (1991). See also: 
P.P. Raychev, R.P. Roussev, and Yu.F. Smirnov, J. Phys. G: Nucl. Part. Phys. {\bf 16}, L137 (1990); 
D. Bonatsos, P.P. Raychev, R.P. Roussev, and Yu.F. Smirnov,
Chem. Phys. Lett. {\bf 175}, 300 (1990); 
B.I. Zhilinski\u\i ~and Yu.F. Smirnov, Soviet J. Nucl. Phys. {\bf 54}, 10 (1991).

\bibitem{Georgieva} 
A. Georgieva and Ts. Dankova, J. Phys. A: Math. Gen. {\bf 27}, 1251 (1994); 
A.I. Georgieva, J.D. Goleminov, M.I. Ivanov, and H.B. Geyer, 
J. Phys. A: Math. Gen. {\bf 32}, 2403 (1999).



\bibitem{Barbier}
R. Barbier, Doctorate thesis, Universit\'e Lyon 1, France, 1995.  
The thesis was partly published in: 
R. Barbier, J. Meyer, and M. Kibler, 
J. Phys. G: Nucl. Part. Phys. {\bf 20}, L13 (1994);
Int. J. Mod. Phys. E {\bf 4}, 385 (1995); 
R. Barbier and M. Kibler, 
A system of interest in spectroscopy: The $qp$-rotor system, in: 
{\it Finite Dimensional Integrable Systems}, eds. A.N. Sissakian and G.S. Pogosyan 
(Joint Institute for Nuclear Research, Dubna, 1995);
On the use of quantum algebras in rotation-vibration spectroscopy, 
in: {\it Modern Group Theoretical Methods in Physics}, eds. J. Bertrand, 
M. Flato, J.-P. Gazeau, D. Sternheimer, and M. Irac-Astaud 
(Kluwer, Dordrecht, 1995); 
Rept. Math. Phys. {\bf 38}, 221 (1996). 

\end{thebibliography}
\end{document}